\documentclass[aps,pre,reprint,english,a4paper,floatfix,showpacs,10pt,superscriptaddress]{revtex4-1}
\usepackage[T1]{fontenc}
\usepackage{ae,aecompl}
\usepackage[english]{babel}
\usepackage{mathtools}
\usepackage[normalem]{ulem}
\usepackage{url}
\usepackage{psfrag,color,pstricks,pst-grad}
\usepackage{amsmath,amsfonts,amssymb,amsbsy}  

\newcommand{\bb}{\mathbf}

\newcommand{\beq}{\begin{equation}}
\newcommand{\eeq}{\end{equation}}
\newcommand{\els}{\varepsilon_{\textrm{ls}}}

\newcommand{\subl}{\textrm{l}}
\newcommand{\subs}{\textrm{s}}

\date{\today}

\begin{document}

\title{Spectral mapping of heat transfer mechanisms at liquid-solid interfaces}
\author{K. S\"a\"askilahti}
\email{kimmo.saaskilahti@aalto.fi}

\author{J. Oksanen}
\author{J. Tulkki}
\affiliation{Engineered Nanosystems group, School of Science, Aalto University, P.O. Box 12200, 00076 Aalto, Finland}
\author{S. Volz}
\email{sebastian.volz@centralesupelec.fr}
\affiliation{\'Ecole Centrale Paris, Grande Voie des Vignes, 92295 Ch\^atenay-Malabry, France}
\affiliation{CNRS, UPR 288 Laboratoire d'Energ\'etique Mol\'eculaire et Macroscopique, Combustion (EM2C), Grande Voie des Vignes, 92295 Ch\^atenay-Malabry, France}

\begin{abstract}
 Thermal transport through liquid-solid interfaces plays an important role in many chemical and biological processes, and better understanding of liquid-solid energy transfer is expected to enable improving the efficiency of thermally driven applications. We determine the spectral distribution of thermal current at liquid-solid interfaces from nonequilibrium molecular dynamics, delivering a detailed picture of the contributions of different vibrational modes to liquid-solid energy transfer. Our results show that surface modes located at the Brillouin zone edge and polarized along the liquid-solid surface normal play a crucial role in liquid-solid energy transfer. Strong liquid-solid adhesion allows also for the coupling of in-plane polarized modes in the solid with the liquid, enhancing the heat transfer rate and enabling efficient energy transfer up to the cut-off frequency of the solid. Our results provide fundamental understanding of the energy transfer mechanisms in liquid-solid systems and enable detailed investigations of energy transfer between, e.g., water and organic molecules. 
\end{abstract}
 \maketitle

 \section{Introduction}
 
Energy transfer across liquid-solid interfaces plays an important role not only in conventional heat exchange applications such as heat pipes \cite{faghri12} but also in emerging nanotechnological applications such as photothermal cancer therapy \cite{lai08}, targeted drug delivery \cite{das07}, solar vapor generation \cite{neumann13}, reversible trapping of biomolecules \cite{huber03}, and nanofluid cooling \cite{keblinski05}. The transfer of vibrational energy between liquids and solids is, however, poorly understood compared to solid-solid interfaces, in part due to the difficulty of experimentally isolating the thermal resistance of the interface  \cite{park12,harikrishna13}. Existing experiments have, however, indicated \cite{wilson02,ge04,ge06,harikrishna13,tian15} that the interfacial conductance of a typical solid-liquid surface is generally close to the conductance of a typical solid-solid interface.

Simulations of the complex vibrational dynamics at liquid-solid interfaces are an invaluable tool in understanding the heat transfer mechanisms. Molecular dynamics (MD) simulations have been used to investigate the effects of, e.g., liquid-solid bonding strength \cite{xue03,barrat03,kim08,giri14}, temperature \cite{ge13}, pressure \cite{pham13}, surface roughness \cite{huang13,zhang14}, surface patterning \cite{issa12}, and surface functionalization \cite{kikugawa09,shenogina09,acharya12,soussi15} on the interfacial conductance. Detailed analysis of heat transfer mechanisms at solid-liquid interfaces is, however, still lacking, mostly due to the nonexistence of simplifying small-displacement approximations allowing for extracting frequency-dependent mode transmission functions as in stiff solid-solid surfaces (see, e.g., Refs. \cite{zhang07,saaskilahti13}). 

In this paper, we gain access to the energy transfer mechanisms at liquid-solid interfaces by utilizing our recently developed method for calculating the spectrally decomposed heat current (SDHC) from nonequilibrium molecular dynamics \cite{saaskilahti14b}. By combining SDHC with the wavevector decomposition \cite{chalopin13,saaskilahti15}, which we have previously developed for analyzing the contributions of different phonon branches on thermal conduction, we provide a thorough picture of the contributions of different vibrational modes on energy transfer at liquid-solid interfaces. In order to highlight the pertinent features in solid-liquid energy transfer instead of addressing material-specific properties, we simulate a generic Lennard-Jones (LJ) system with the potential parameters of liquid argon and ''stiff'' argon as in numerous earlier works \cite{barrat03,xue03,xue04,kim08,caplan14,giri14}. The LJ potential accounts for the pertinent features of interatomic interactions (repulsion and attraction at small and large distances, respectively) and also allows for simulating large systems efficiently.

\section{Theory and methods}

\subsection{Spectral heat flux}
Spectral heat current at the liquid-solid interface is determined by monitoring the force-velocity correlations at the interface \cite{saaskilahti14b}. The starting point for the analysis is the total heat flux flowing across the interface from the liquid (l) to the solid (s) \cite{hardy63,saaskilahti12}, 
\begin{alignat}{2}
 Q_{\subl \to \subs} &  = \frac{1}{2A} \sum_{j \in \subl} \sum_{i\in \subs}  \langle \bb{F}_{ij} \cdot (\bb{v}_i +\bb{v}_j) \rangle \\
  &= \frac{1}{A} \sum_{j \in \subl} \sum_{i\in \subs} \langle \bb{F}_{ij} \cdot \bb{v}_i \rangle \label{eq:Q_2}.
\end{alignat}
Here $A$ is the interface area, the interparticle force $\bb{F}_{ij}$ is defined as the derivative of the interatomic interaction potential $V_{ij}$ with respect to particle position $\bb{x}_i$,
\begin{equation}
 \bb{F}_{ij} = -\frac{\partial V_{ij}}{\partial \mathbf{x}_i},
\end{equation}
and the identity \cite{dhar08} $\langle \bb{F}_{ij} \cdot \bb{v}_j \rangle=\langle \bb{F}_{ij} \cdot \bb{v}_i \rangle$ was used in Eq. \eqref{eq:Q_2}. Following the derivation of Ref. \cite{saaskilahti14b}, the heat flux $Q_{\subl \to \subs}$ can be decomposed spectrally as
\begin{equation}
 Q_{\subl \to \subs}  = \int_0^{\infty} \frac{d\omega}{2\pi} q_{\subl \to \subs}(\omega),
\end{equation}
where $\omega$ is the angular frequency and the spectral heat flux $q_{\subl \to \subs}(\omega)$ is given by the expression
\begin{equation}
 q_{\subl \to \subs}(\omega) = \frac{2}{A}\textrm{Re}  \sum_{j \in \subl} \sum_{i\in \subs} \int_{-\infty}^{\infty} d\tau e^{i\omega \tau} \langle \bb{F}_{ij}(\tau) \cdot \bb{v}_i(0)  \rangle. \label{eq:qom}
\end{equation}
Here $\tau$ is the correlation time between forces and velocities and the correlation function $\langle \bb{F}_{ij}(t_1) \cdot \bb{v}_i(t_2)  \rangle$ only depends on the time-difference $t_1-t_2$ due to the assumed steady state. 

\begin{figure}
 \begin{center}
  \includegraphics[width=8.6cm]{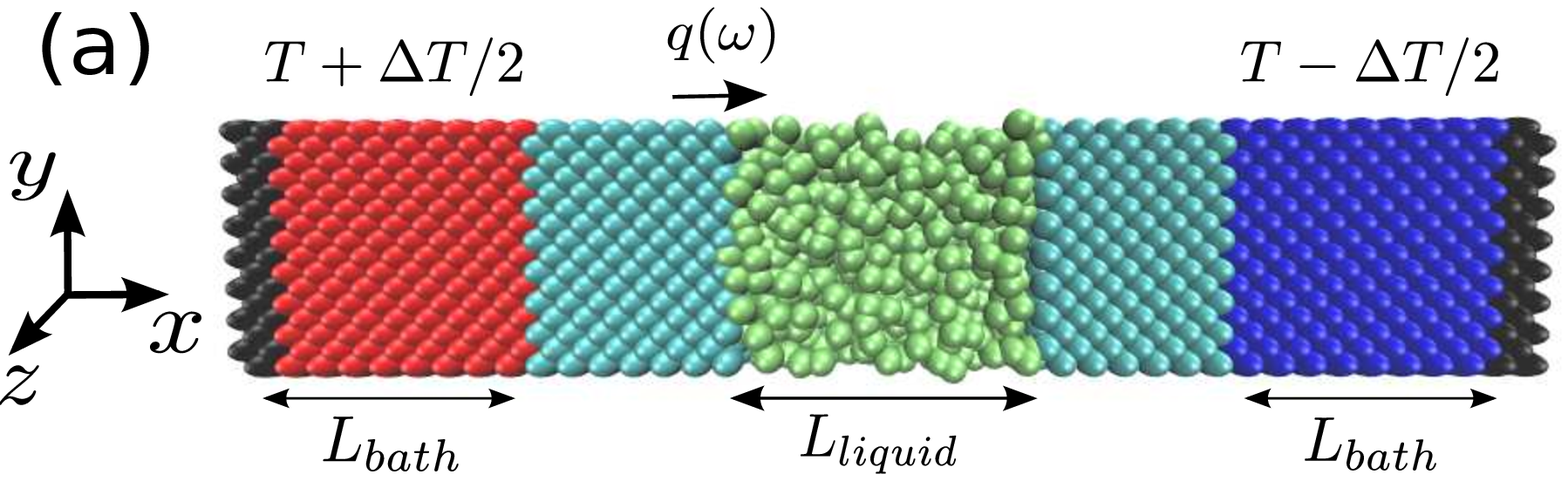}
  \includegraphics[width=8.6cm]{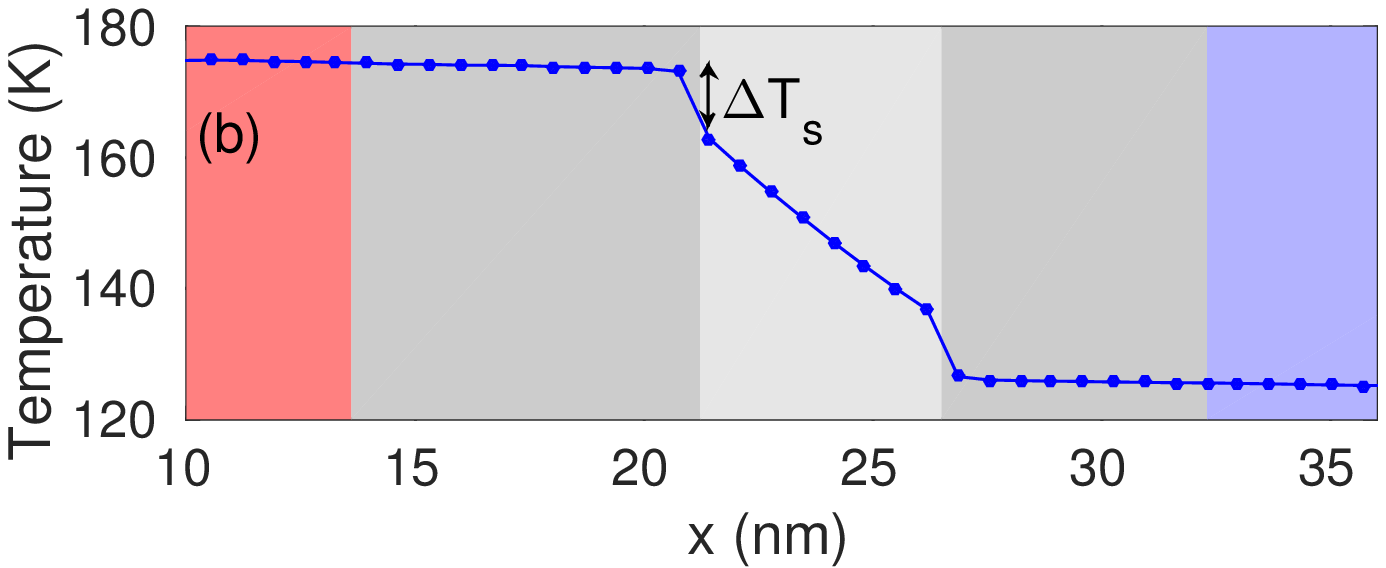}
 \caption{(Color online) (a) Schematic illustration of the studied system: Lennard-Jones liquid sandwiched between two Lennard-Jones solids arranged in a face-centered cubic lattice. The spectral heat current $q_{\subl\to\subs}(\omega)$ is calculated at the solid-liquid interfaces by monitoring the force and velocity trajectories of the atoms located within the potential cut-off distance from the liquid. (b) Typical temperature profile in the non-equilibrium simulation. The spectral conductance is defined as $g(\omega)=|q_{\subl\to\subs}(\omega)|/\Delta T_s$, where $\Delta T_s$ is the temperature jump at the surface.}
 \label{fig:geom}
 \end{center}
\end{figure}

\subsection{Simulation setup}

We investigate the spectral heat flux in the simulation system depicted in Fig. \ref{fig:geom}. LJ liquid of length $L_{liquid}$ is sandwiched between two LJ solids consisting of $40\times 8 \times 8$ unit cells in a face-centered cubic lattice oriented so that the [100] crystal direction is along the $x$-axis. The LJ interatomic interaction energy is \cite{allentildesley}
\begin{equation}
 V_{ij}(r) = 4\varepsilon_{ij} \left[\left(\frac{\sigma}{r}\right)^{12}- \left(\frac{\sigma}{r}\right)^6 \right], 
\end{equation}
where the particle sets $i,j\in \{l,s\}$ label either liquid ($i\in l$ or $j\in l$) or solid ($i \in s$ or $j\in s$) and $r$ is the interparticle distance. We choose the potential and mass parameters of the liquid to correspond to liquid argon \cite{rahman64}: $\sigma=3.4$ \AA, $\varepsilon_{ll}=1.67\times 10^{-21}$ J, and atomic mass $m=39.948$ amu. For the solid parts, $m$ and $\sigma$ are unchanged but the interaction strength is chosen as $\varepsilon_{ss}=10\varepsilon_{ll}$ as in earlier simulations \cite{xue03,xue04,giri14}. The liquid-solid interaction parameter is set to $\varepsilon_{ls}=\varepsilon_{ll}$ unless noted otherwise and the interaction cut-off radius is $r_c=2.5\sigma$. The length of the liquid region is chosen as $L_{liquid} \approx 5.5$ nm. 

To impose nonequilibrium, atoms located in regions of length $L_{bath}=13.6$ nm are coupled to Langevin heat baths at temperatures $T+\Delta T/2$ and $T-\Delta T/2$ in the left and right solids, respectively. We use the Langevin bath relaxation time $\tau=2.14$ ps, corresponding to weak bath coupling \cite{saaskilahti14b}. Simulations are performed at the mean temperature $T=150$ K and with the temperature bias $\Delta T=50$ K. System pressure is set to $P=50$ MPa by controlling the system volume with a barostat in an initial equilibrium simulation. In this equilibrium run, periodic boundary conditions are employed in all three spatial directions. The simulations were performed at the average temperature of $T=150$ K, which is close to the value of $T=170$ K recently used in Ref. \cite{giri14}. At $T=150$ K, relatively high pressure is needed to ensure that the pair distribution function of the middle part of the system exhibits the oscillations expected in the liquid phase \cite{allentildesley}, and we choose the value $P=50$ MPa for our simulations. After the equilibration, periodic boundary conditions are imposed only in the $y$- and $z-$directions and atoms located in the two left- and right-most unit cells are fixed to avoid atomic sublimation at the system boundaries. Atomic equations of motion are integrated using LAMMPS simulation software \cite{plimpton95} with the timestep $4.28$ fs. 

Atomic velocities and forces are collected for the evaluation of the spectral heat flux \eqref{eq:qom} during a nonequilibrium collection run of $10^7$ steps. Details of the evaluation of Eq. \eqref{eq:qom} from the MD trajectories are discussed in Appendix \ref{app:A}. A typical nonequilibrium temperature profile observed in the simulation is shown in Fig. \ref{fig:geom}(b). As expected, the temperature profile exhibits a temperature jump $\Delta T_s$ at the interface due to the non-zero thermal resistance. In the results below, we normalize the calculated spectral heat flux by the temperature jump to extract the spectral interface conductance as $g(\omega)=|q_{\subl\to\subs}(\omega)|/\Delta T_s$. Here the absolute value accounts for the fact that at the left interface, heat flows from the solid to the liquid. Temperature jump $\Delta T_s$ at the surface is determined by extrapolating linearly fitted temperature profiles in the solid and the liquid to the interface and calculating the difference \cite{pham13}. Note that the precise values of $\Delta T_s$ are not critically important in our analysis, because we focus on the \textit{spectral distribution} of conductance instead of absolute values. The effects of structural parameters, pressure, and solid-liquid mass ratio on the spectral conductance are discussed in the Appendix.

\begin{figure}
 \begin{center}
  \includegraphics[width=8.6cm]{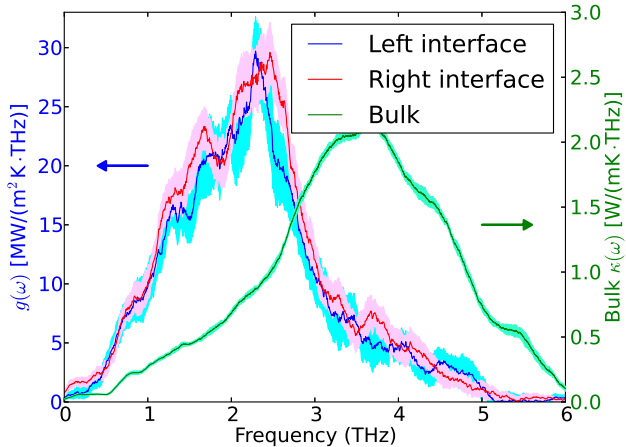}
 \caption{(Color online) Spectral interface conductance $g(\omega)=|q(\omega)|/\Delta T_s$ at the left and right liquid-solid interfaces versus frequency. For reference, also the spectrally decomposed bulk conductivity of the solid is shown.}
 \label{fig:1}
 \end{center}
\end{figure}

\section{Numerical results}

Figure \ref{fig:1} shows the spectral interface conductance $g(\omega)=|q_{\subl \to \subs}(\omega)|/\Delta T_s$ versus frequency $f=\omega/(2\pi)$ at the left and right interfaces. Conductances calculated at the left and right interfaces agree closely, suggesting that the conductance is relatively insensitive to the temperature difference between the left and right interfaces. In the analysis below, we therefore focus on the spectral conductance at the left interface. 

The spectral interface conductance of Fig. \ref{fig:1} has a maximum at $f\approx 2.4$ THz, indicating the dominant contribution of vibrations with frequencies around this value to the interfacial heat transfer. The position of the peak differs from the maximum position in the bulk spectral conductivity $\kappa(\omega)=q(\omega)/|dT/dx|$ calculated for the two solids in direct contact without the liquid ($dT/dx$ is the thermal gradient in the bulk). This shift in the position of maximum is similar to the shift in the density of states at a free surface compared to the one of the bulk \cite{allen71}. Figure \ref{fig:1} also shows that the highest-frequency modes in the bulk at frequencies $f\gtrsim 5$ THz do not contribute to liquid-solid energy transfer, indicating the decoupling of high-frequency vibrations in the solid from the vibrations in the liquid.

\begin{figure}[t]
 \begin{center}
  \includegraphics[width=4.2cm]{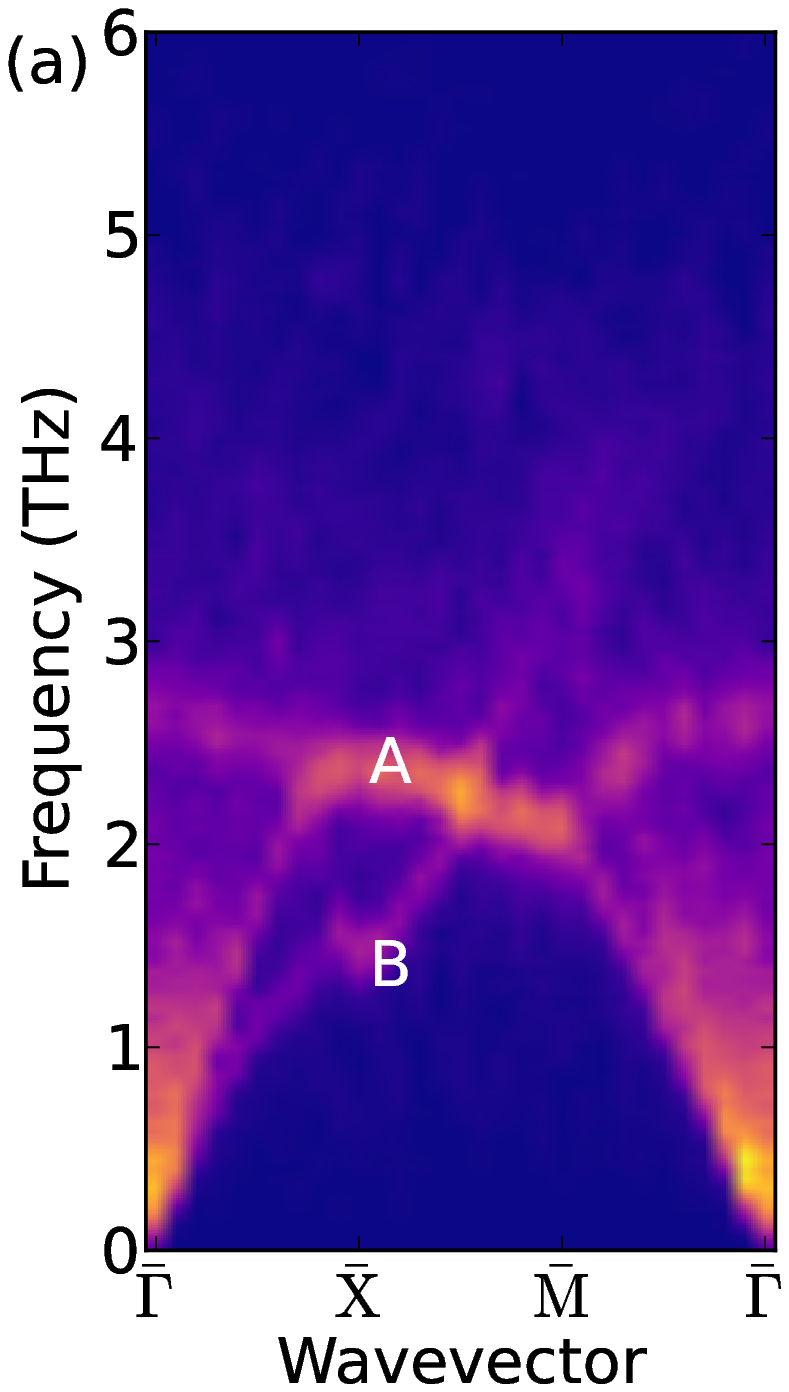}
  \includegraphics[width=4.2cm]{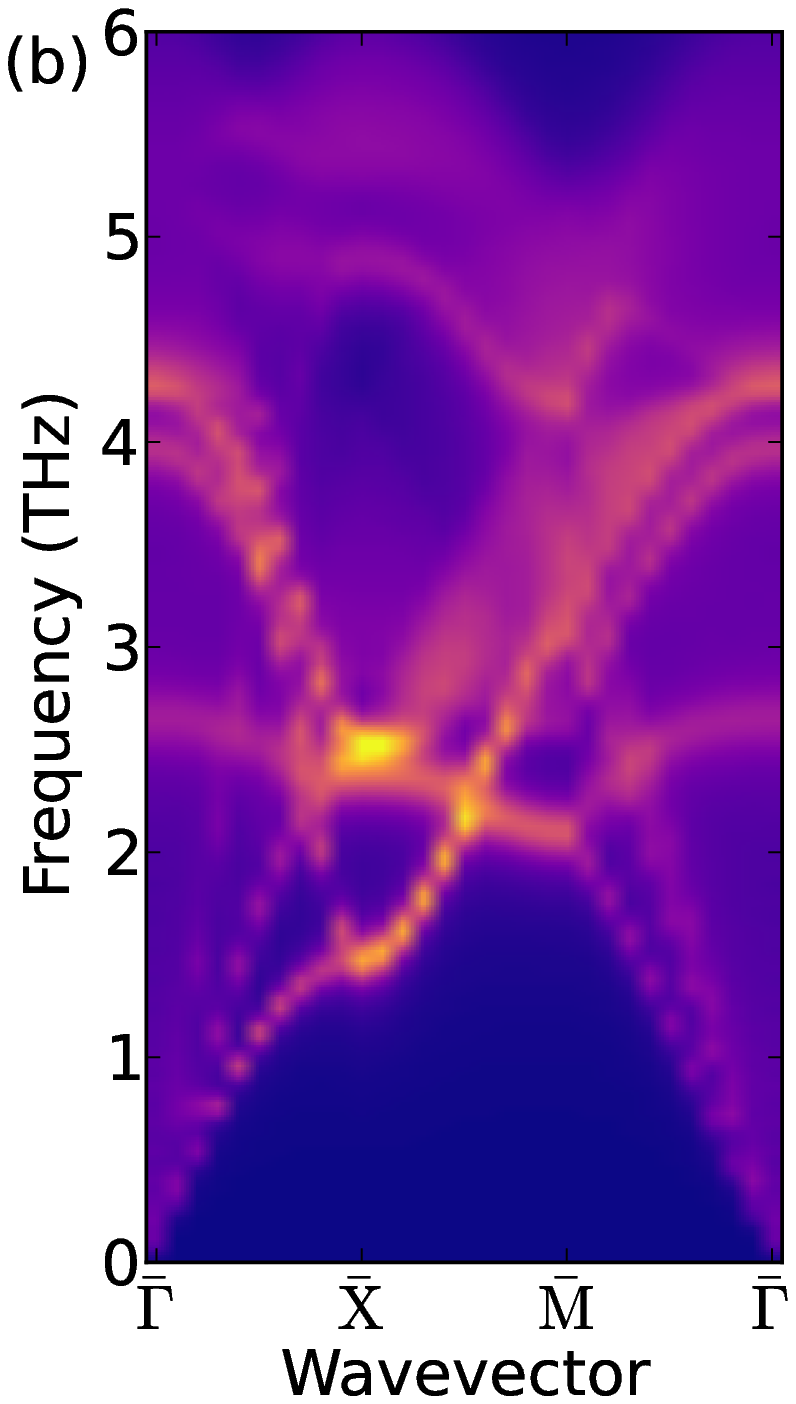} \\
  \includegraphics[width=4.2cm]{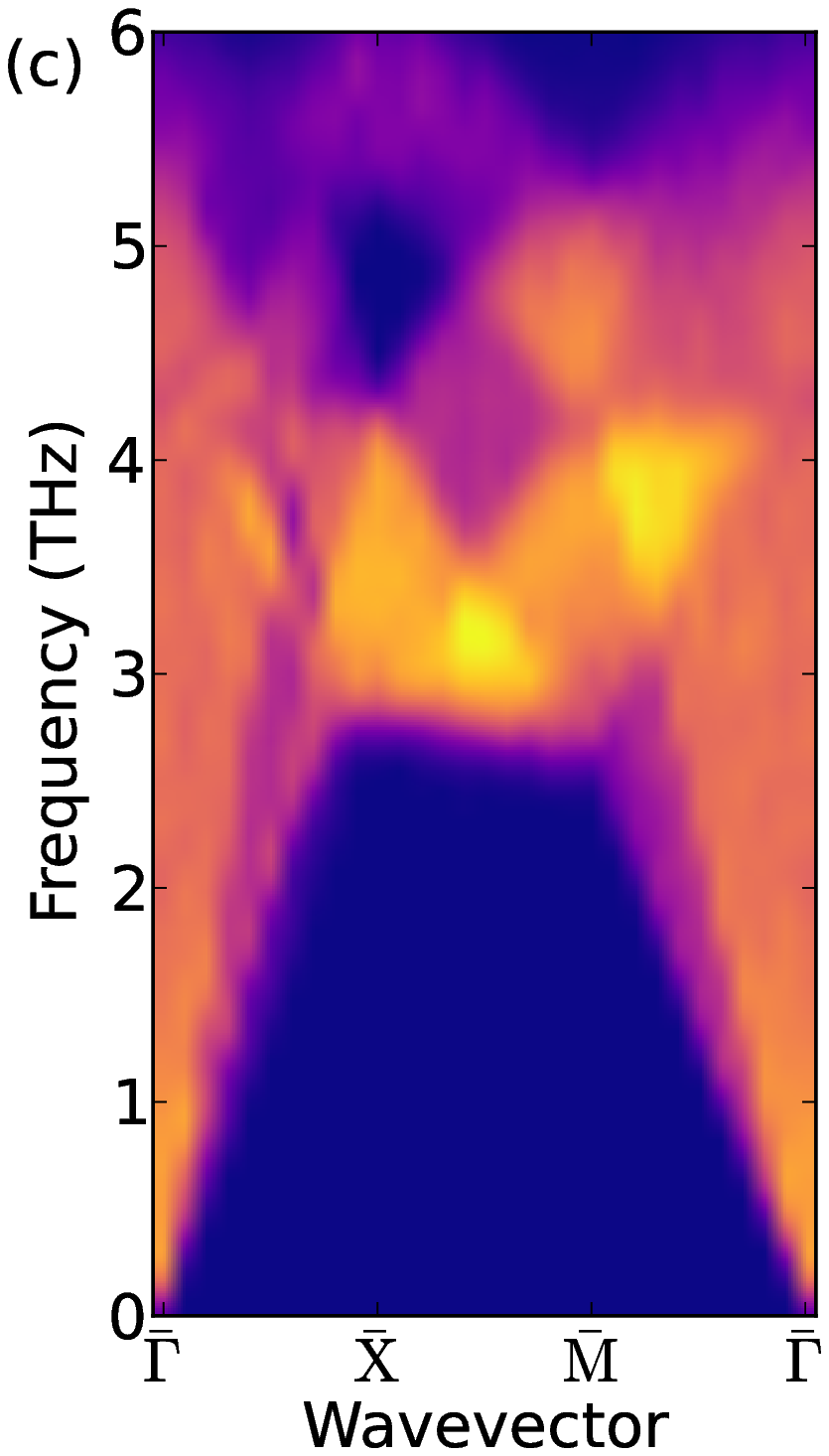}
  \includegraphics[width=4.2cm]{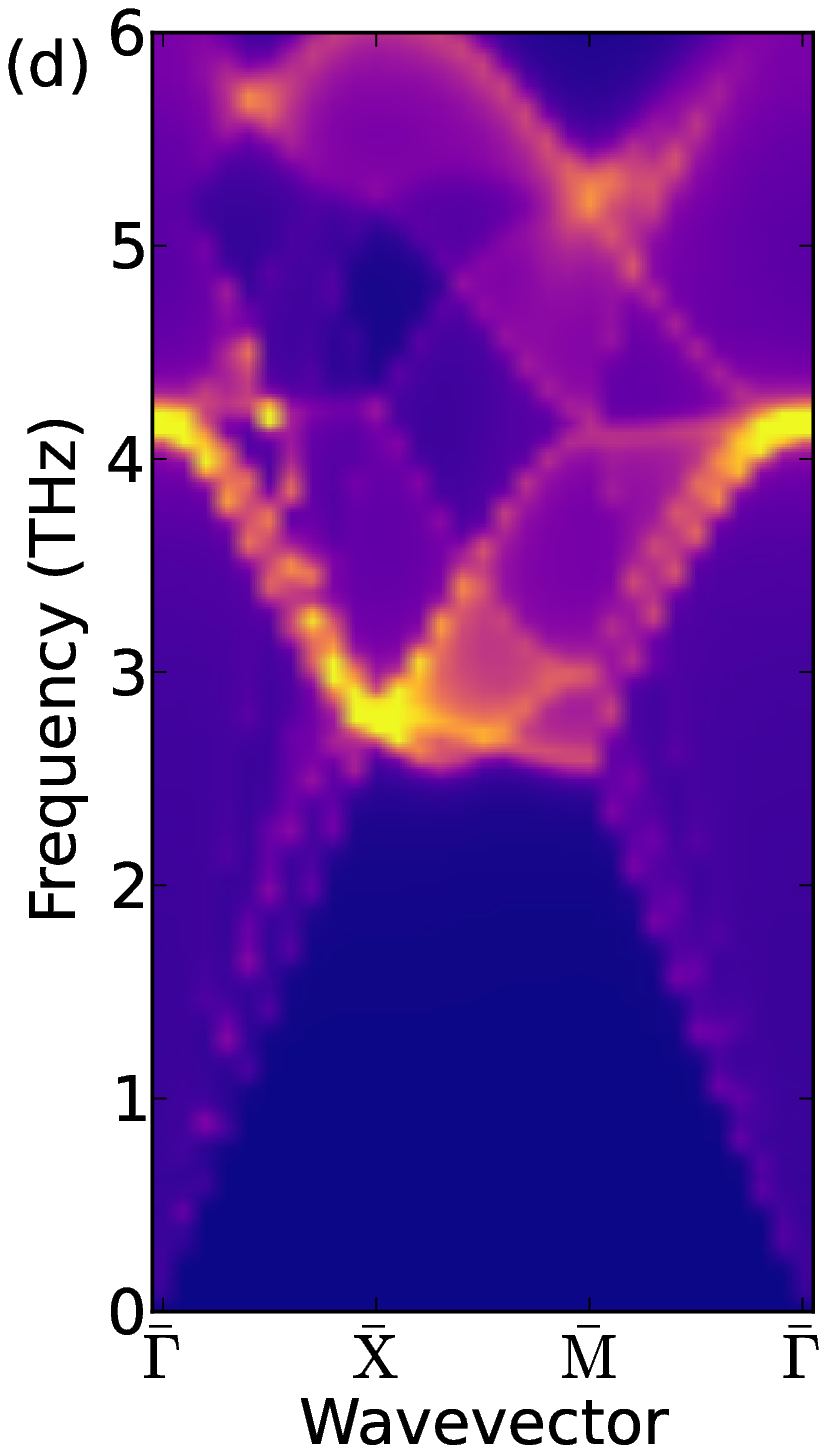}
 \caption{(Color online) (a) Wavevector-decomposed  spectral heat flux and (b) interfacial density of states versus wavevector and frequency at the solid-liquid interface (arbitrary units, brighter colors correspond to larger values). (c) Wavevector-decomposed  spectral heat flux and (d) interfacial density of states in the bulk solid. The surface mode labeled as A in (a) and located on the Brillouin zone boundary $\bar{\textrm{X}}$--$\bar{\textrm{M}}$ contributes strongly to the interfacial conductance at $f\approx 2.3$ THz. Surface mode labeled B has, on the other hand, smaller effect on energy transfer due to its in-plane polarization. Figures have been calculated for a $20\times 20$ cross-section to improve the wavevector resolution compared to the $8\times 8$ cross-section used in other calculations.}
 \label{fig:2}
 \end{center}
\end{figure}

To better understand the shape of the spectral distribution of Fig. \ref{fig:1}, we take advantage of the translational invariance along the $y$- and $z$-directions and decompose Eq. \eqref{eq:qom} into in-plane wavevector contributions as
\begin{equation}
 q(\omega) = \sum_{\bb{q}_{\parallel} \in SBZ} q(\omega;\bb{q}_{\parallel}),
\end{equation}
where $\bb{q}_{\parallel}=(q_y,q_z)$ is the in-plane wavevector and SBZ is the two-dimensional Brillouin zone of the (100) surface \cite{allen71}. The decomposition is derived  by substituting the two-dimensional discrete Fourier transforms of forces $\bb{F}_i$ and velocities $\bb{v}_i$ in Eq. \eqref{eq:qom} and performing the sum over unit cells in $y$- and $z$-directions to eliminate one of the wavevectors \cite{chalopin13,saaskilahti15}. 

Figure \ref{fig:2} shows the wavevector-decomposed spectral current at (a) the liquid-solid interface and (c) in bulk solid along different directions in the two-dimensional Brillouin zone. The wavevectors labeled as $\bar{\Gamma}$, $\bar{\textrm{X}}$ and $\bar{\textrm{M}}$ correspond to the high-symmetry points in the ($q_y$,$q_z$)-plane as in Ref. \cite{allen71}. It can be seen from Fig. \ref{fig:2}(a) that modes at the Brillouin zone boundary $\bar{\textrm{X}}$--$\bar{\textrm{M}}$ contribute strongly to the peak in the spectral conductance at $f\approx 2.3$ THz (branch labeled A). By comparing the vibrational density of states at the surface of the solid [Fig. \ref{fig:2}(b)] to the bulk density of states of the solid [Fig. \ref{fig:2}(d)], these modes at the zone boundary at $f\approx 2.3$ THz can be identified as surface modes due to their absence in the bulk \cite{allen71} (note that the lowest-frequency modes at the interface in the $\bar{\textrm{X}}$--$\bar{\textrm{M}}$ line have a slightly smaller frequency than the lowest-frequency bulk modes). The surface density of states of Fig. \ref{fig:2}(b) also shows a low-frequency surface mode in the $\bar{\Gamma}$--$\bar{\textrm{X}}$ and $\bar{\textrm{X}}$--$\bar{\textrm{M}}$ directions [labeled B in Fig. \ref{fig:2}(a)]. The small contribution of this surface mode on the spectral heat flux of Fig. \ref{fig:2}(a) can be understood by the fact that the mode is mostly polarized in the in-plane direction \cite{allen71} and cannot therefore couple to the liquid, as discussed in more detail below. The branch labeled A is, on the other hand, primarily polarized in the out-of-plane direction, allowing for efficient coupling between the solid and the liquid. 

In addition, Fig. \ref{fig:2}(a) shows that the acoustic modes at low frequencies ($\lesssim 1$ THz) and near the zone center $\bar{\Gamma}$ couple efficiently with the liquid. Their total contribution to the spectral conductance of Fig. \ref{fig:1} is, however, small because they occupy only a small region of the Brillouin zone in the ($q_y$,$q_z$)-plane.

\begin{figure}
 \begin{center}
  \includegraphics[width=8.6cm]{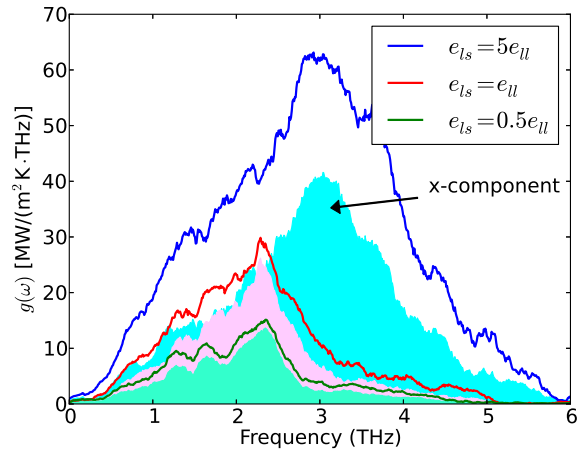}
 \caption{(Color online) Spectral conductance at the left liquid-solid interface for various liquid-solid interaction parameters $\varepsilon_{\textrm{ls}}$. The shaded regions denote the contribution of forces and velocities oriented along the $x$-coordinate (normal to the liquid-solid surface) to the total conductance [Eq. \eqref{eq:qom_x}].}
  \label{fig:els}
 \end{center}
\end{figure}

Parameter values $\els>1$ and $\els<1$ correspond, respectively, to ''wetting'' and ''non-wetting'' liquids \cite{xue03}. In the former case, the liquid tends to form a wetting layer on the solid surface \cite{liang11}, enhancing the heat transfer rate \cite{ge06}. It has been suggested \cite{caplan14} that the enhancement arises from the improved coupling of in-plane vibrations at the solid surface with the liquid, but this has not been transparently demonstrated. To understand the role of interfacial bonding strength on heat transfer, we show in Fig. \ref{fig:els} the spectral conductance for various liquid-solid interaction parameters $\varepsilon_{\textrm{ls}}$. To analyze the roles of in-plane and out-of-plane vibrations (in and perpendicular to the $yz$-plane, respectively), we also show the $x$-coordinate (out-of-plane) contribution
\begin{equation}
 g^x(\omega) = -\frac{2}{A\Delta T_s}\textrm{Re}  \sum_{j \in \subl} \sum_{i\in \subs} \int_{-\infty}^{\infty} d\tau e^{i\omega \tau} \langle {F}_{ij}^x(\tau) {v}^x_i(0)  \rangle. \label{eq:qom_x}
\end{equation}
on the spectral conductance as shaded regions.

In the case of weak liquid-solid bonding ($\els=0.5\varepsilon_{ll}$), which corresponds to a hydrophobic surface, the contribution of in-plane vibrations to the spectral conductance of Fig. \ref{fig:els} is negligible. In this case, the liquid atoms satisfy a microscopic free slip boundary condition, allowing the liquid atoms to flow freely along the surface, decoupled from the in-plane vibrations of the solid \cite{caplan14}. It is, however, important to note that in-plane vibrations can be intimately coupled to out-of-plane motion as in, e.g., Rayleigh waves.

In the case of a strong liquid-solid bonding ($\els=5\varepsilon_{ll}$), on the other hand, the spectral conductance increases at all frequencies. At frequencies $f\lesssim 2$ THz, the increase originates mostly from the increased contribution of in-plane vibrations: strong bonding essentially renders a no-slip boundary condition \cite{caplan14} at the surface, forcing the velocities of liquid atoms to follow the velocities of solid atoms and allowing for energy transfer. At frequencies $f\gtrsim 2$ THz, the contribution of in-plane vibrations remains large, but also vibrations along the surface normal ($x$-axis) benefit from the strong coupling and contribute more to the conductance: even the highest-frequency modes with $f\gtrsim 5$ THz can couple to the liquid. The large contribution of high-frequency modes to the conductance in the case of strong coupling is in contrast with the predictions of Ref. \cite{giri14}, where the analysis of the interfacial density of states in different cases suggested that the increase in the conductance is likely to originate from low-frequency modes. The difference in the conclusion is expected to arise because the interfacial density of states is only indirectly related to the thermal conductivity and may therefore fail to reflect all the pertinent features of the spectral heat current. In contrast, our spectral decomposition method directly decomposes the thermal current as described by MD into spectral domain and therefore is expected to avoid potential complications in interpreting the spectral form of the heat current. The hypotheses of Refs. \cite{caplan14,giri14} of the importance of in-plane vibrations in the case of strong solid-liquid bonding is, however, validated by our results. 

It is important to note that in the strongly bonded case ($\els=5\varepsilon_{ll}$), the liquid atoms closest to the solid adsorb to the solid surface, enabling the transfer of high-frequency vibrations between the solid and this wetting layer. We do not expect, however, that these high-frequency vibrations are able to carry energy deep into the liquid. Instead, they are most likely transformed into lower-frequency vibrations carrying the energy. In future, it could be fruitful to investigate the validity of this hypothesis by calculating the spectral decomposition of heat current \textit{inside} the liquid.

\section{Summary}

In summary, we have determined the spectral distribution of thermal conductance of liquid-solid interfaces, providing fundamental microscopic understanding of the heat transport mechanisms. Energy transfer across the interface was shown to be dominated by the surface modes located at the Brillouin zone boundary and polarized along the surface normal (out-of-plane). When the liquid-solid bonding is weak, only out-of-plane vibrations along the surface normal contribute to energy transfer, but strong interfacial bonding (hydrophilicity) allows also for the coupling of transverse vibrations in the solid with the liquid. Strong bonding also enables heat transfer at very high frequencies, extending up to the cut-off frequency of the solid. The presented results and methods lay ground for future investigations of heat transfer mechanisms between prevalent liquids such as water and complex materials such as functionalized surfaces, nanoparticles, and biomolecules. 

\section{Acknowledgements} 

The computational resources were provided by the Finnish IT Center for Science and Aalto Science-IT project. The work was partially funded by the Aalto Energy Efficiency Research Programme (AEF) and the Academy of Finland.

\appendix

\section{Evaluation of the heat current spectrum}
\label{app:A}

In the results section of the main manuscript, we report the spectral heat flux distributions $q(\omega)=|q_{\textrm{l}\to\textrm{s}}(\omega)|$ at both left and right liquid-solid interfaces. As discussed in Ref. \cite{saaskilahti15}, the spectral flux can be calculated in terms of the discrete Fourier transforms of the force and velocity trajectories of the atoms in the left solid as
\begin{equation}
  q_{\textrm{l} \to \textrm{s}}(\omega) = \frac{2}{AM\Delta t_{s}}  \textrm{Re} \sum_{j \in \textrm{l}} \sum_{i\in \textrm{l}}\langle \hat{\bb{F}}_{i}^{liquid}(\omega) \cdot \hat{\bb{v}}_i(\omega)^*  \rangle, \label{eq:q_wiener}
\end{equation}
where
\begin{equation}
 \bb{F}_{i}^{liquid} = \sum_{j\in L} \bb{F}_{ij}
\end{equation}
is the total force exerted by the liquid on the solid atom $i$ and the Fourier transforms are defined for, e.g., the velocities as
\begin{equation}
 \hat{\bb{v}}_i(\omega_l) = \Delta t_s \sum_{k=1}^M e^{i\omega_l k \Delta t_s} \bb{v}_i(k\Delta t_s).
\end{equation}
Here $\Delta t_s$ is the sampling interval, $M$ is the number of samples, and the discrete angular frequencies are $\omega_l=2\pi l/(M\Delta t_s)$. By dividing the simulation run into blocks of size $M$, determining the spectral current for each block and calculating the inter-block variances, we can estimate the statistical error in the spectral currents. We use $\Delta t_s=10\Delta t=42.8$ fs and $M=50000$ in our calculations. It is important to note that only the atoms located within the potential cut-off distance $r_c$ from the liquid contribute to Eq. \eqref{eq:q_wiener}. Finally, the sharply fluctuating spectra obtained from Eq. \eqref{eq:q_wiener} require spectral smoothing, which is carried out with a rectangular window of width $\Delta \omega=2 \pi \times 0.23$ THz.

\section{Effect of liquid length and cross-section size}
\label{app:B}
\begin{figure}
 \begin{center} 
  \includegraphics[width=.99\columnwidth]{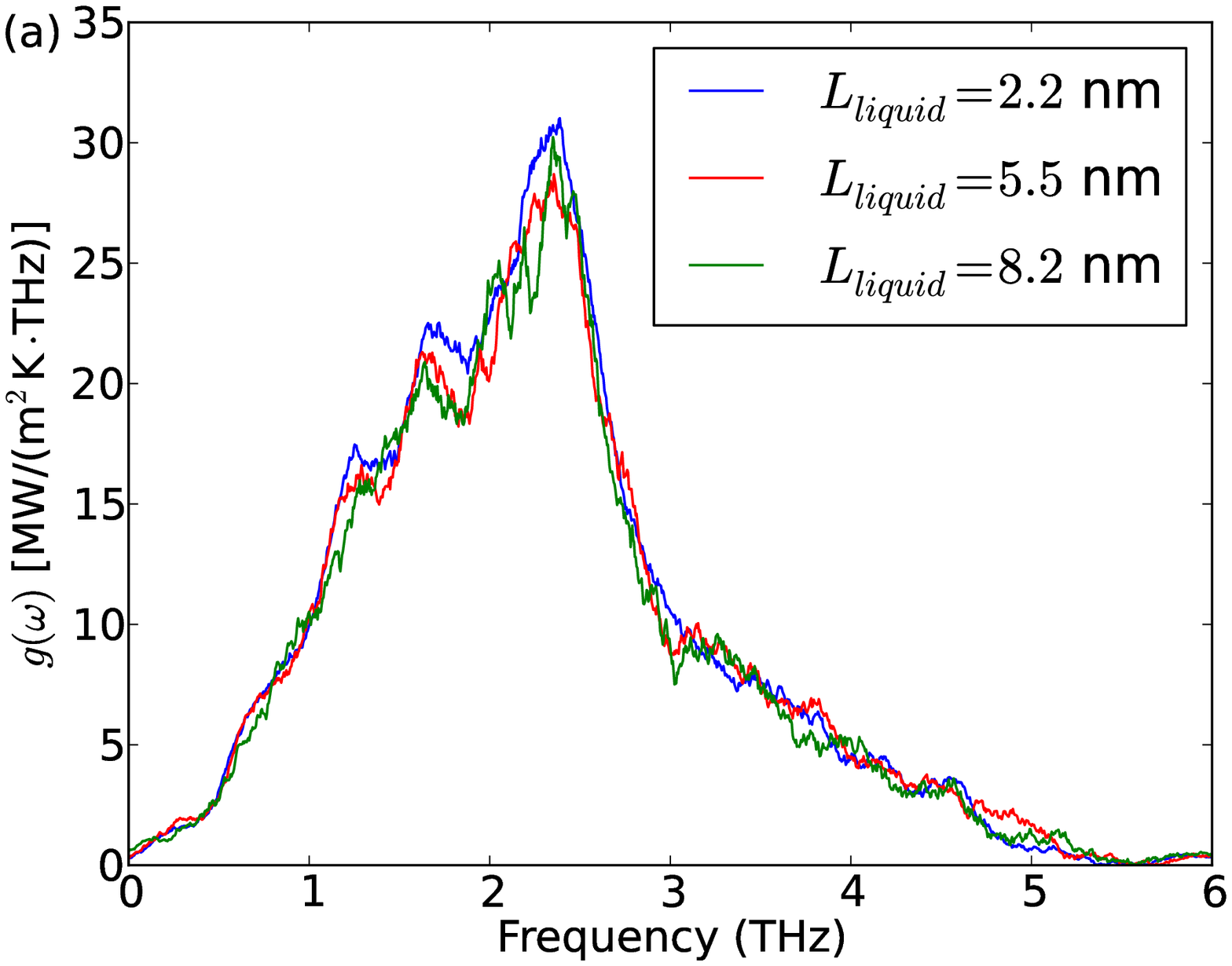}
  \includegraphics[width=.99\columnwidth]{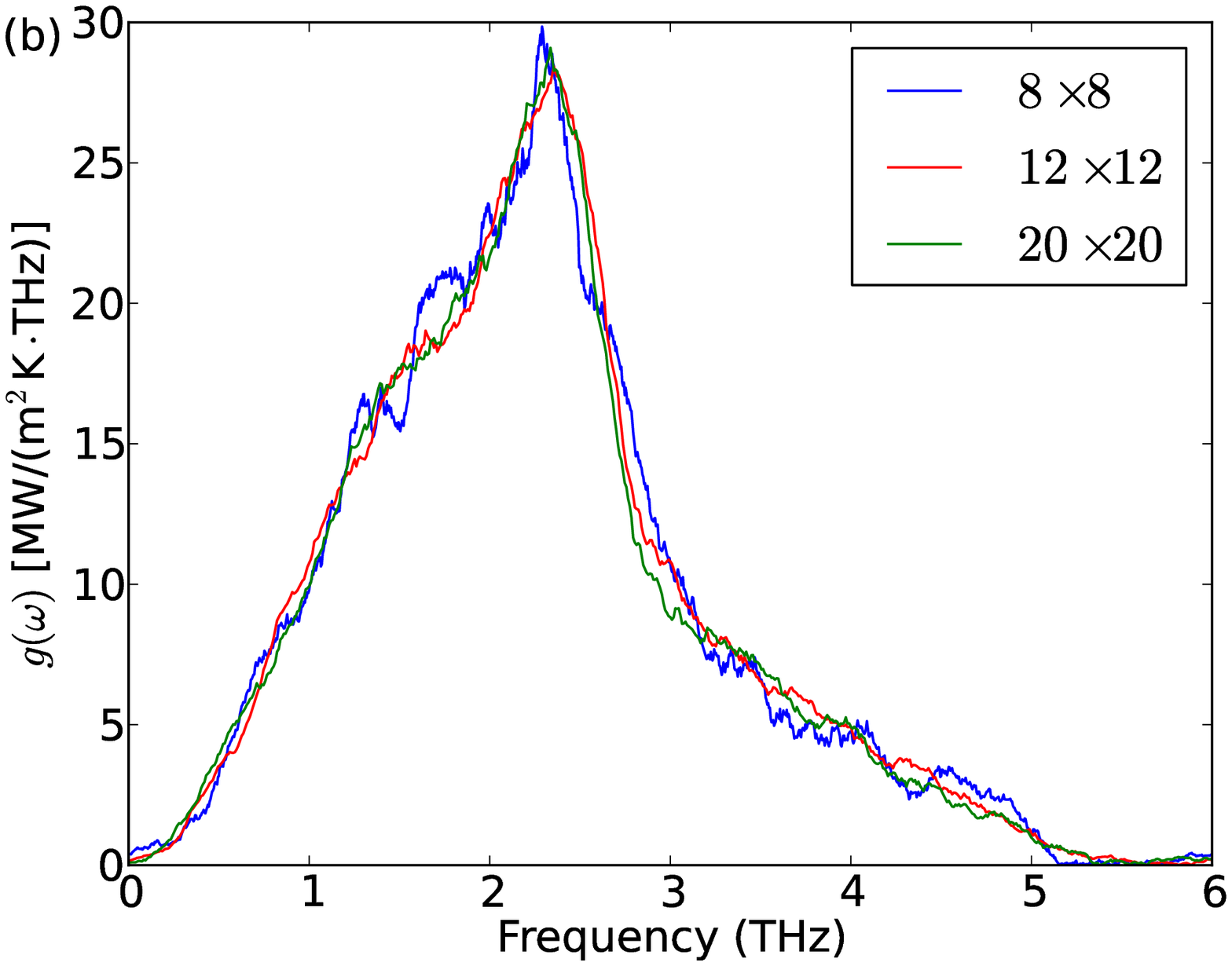}
 \caption{(a) Spectral conductance $g(\omega)$ at the left interface for various lengths of the liquid at $T=150$ K and $P=50$ MPa. (b) Spectral conductance for various numbers of unit cells in the cross-section.} 
 \label{fig:L}
 \end{center}
\end{figure}

The spectral conductance $g(\omega)$ shown in Fig. 2 of the main manuscript can only be considered an inherent property of the surface, if it is insensitive to the liquid length $L_{liquid}$. In Fig. \ref{fig:L}(a), we therefore plot the spectral heat flux and the conductance (obtained by dividing by $\Delta T^{l-s}$ for various values of $L_{liquid}$. It can be seen in Fig. \ref{fig:L}(b) that except for the shortest liquid considered, $L=2.2$ nm, the conductance is essentially independent of the liquid length. This independence is closely related to the diffusivity of heat transfer in the liquid: if the mean free paths of heat carriers exceeded the length of the liquid, local conductance could not be unambiguously defined (which is the case in bulk superlattices).

Figure \ref{fig:L}(b) shows the spectral conductance for various cross-section sizes. Here, for example, $8\times 8$ refers to eight unit cells in $y$- and $z$-direction. The results of Fig. \ref{fig:L}(b) show that the conductance is not sensitive to the cross-section size and, therefore, the results shown for $8\times 8$ cross-section in the main manuscript do not suffer from finite-size effects.

\section{Spectral heat current in the bulk}
\label{app:C}
\begin{figure}
 \begin{center}
  \includegraphics[width=.99\columnwidth]{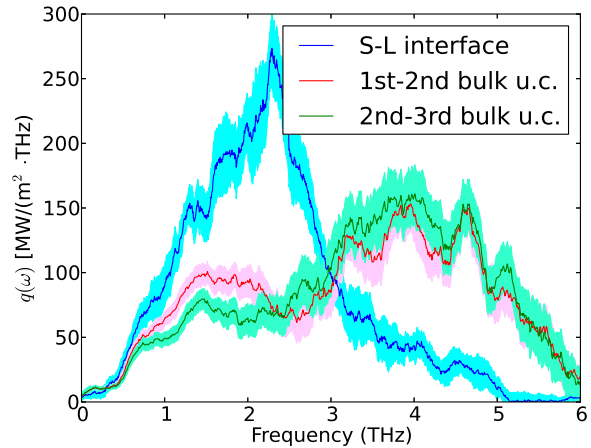} 
\caption{Spectral heat current $q(\omega)$ inside the solid in the immediate vicinity of the interface, compared to the current at the solid-liquid interface. In the labeling of heat currents in the solid, the first unit cell (u.c.) refers to the two solid monolayers closest to the liquid, second unit cell refers to the third and fourth monolayers away from the interface, etc. }
 \label{fig:shc_bulk}
 \end{center}
\end{figure}

Figure \ref{fig:shc_bulk} compares the spectral heat current at the liquid-solid interface to the heat currents inside the solid in the vicinity of the interface. The calculation of spectral currents in the solid has been described earlier in Ref. \cite{saaskilahti14b}. Figure shows that the frequency distribution of heat current is very different in the solid and the liquid. In the solid, the spectral distribution is similar to the bulk distribution in the absence of liquid (Fig. 2 in the main manuscript), with a peak at $\sim 4$ THz. This analysis shows that the localized surface modes, which are strongly coupled with the liquid and therefore contribute to solid--liquid heat transfer at $\lesssim 2.5$ THz, are able to couple also with the propagating bulk modes and therefore mediate energy transfer.

\section{Effect of pressure}
\label{app:D}

\begin{figure}
 \begin{center}
  \includegraphics[width=.99\columnwidth]{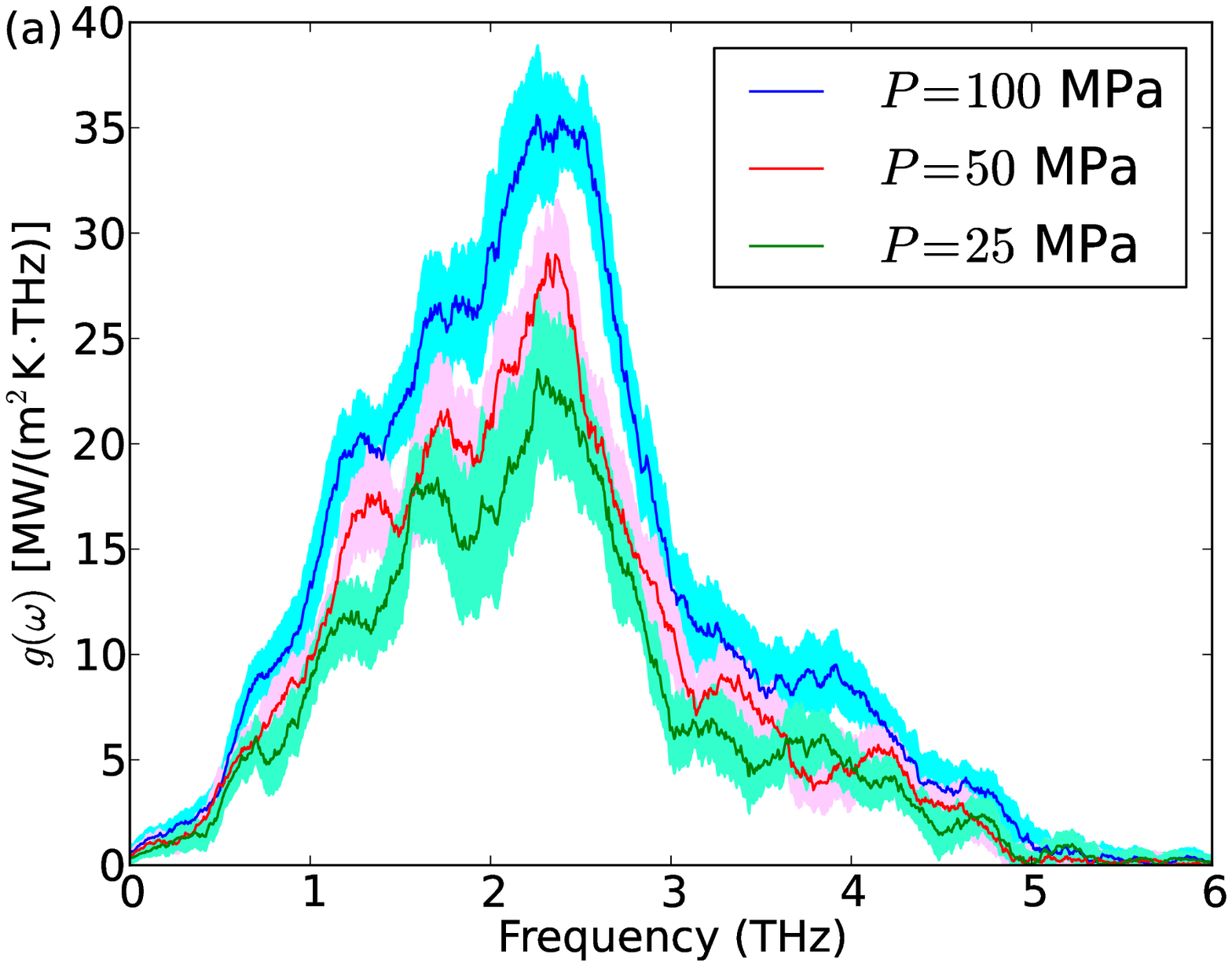}
   \includegraphics[width=.99\columnwidth]{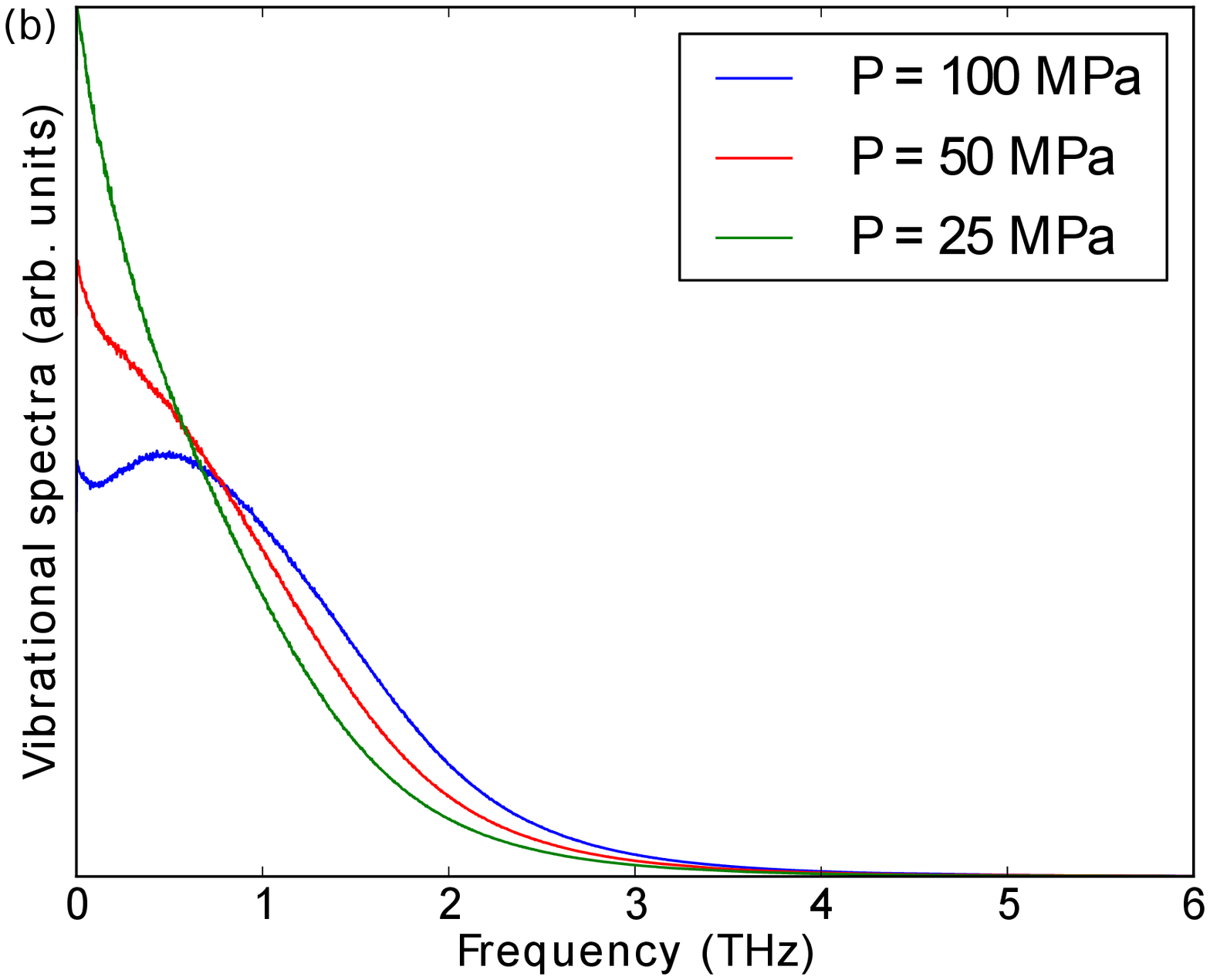}
 \caption{(a) Spectral conductance $g(\omega)$ at the left liquid-solid interface for different pressures $P$. (b) Vibrational spectrum in the liquid in the three cases.}
 \label{fig:P}
 \end{center}
\end{figure}

Exerting pressure on the solid-liquid system is known to boost the interfacial conductance at hydrophilic interfaces \cite{pham13}, because the increasing pressure enables closer contact between the liquid and the solid. To investigate which frequencies play the main role in the enhancement, we plot in  Fig. \ref{fig:P}(a) the spectral conductance at pressures $P=25$ MPa, $P=50$ MPa (the case of Fig. 2 in the manuscript) and $P=100$ MPa for $\varepsilon_{ls}=1$. Figure \ref{fig:P}(a) shows the contribution of low-frequency vibrations at $\lesssim 1$ THz to the spectral heat current depends only weakly on pressure. In the frequency range from 1 to 5 THz, however, increasing the pressure increases the spectral conductance, and the enhancement is especially notable at $\gtrsim 2$ THz. The calculation of Fig. \ref{fig:P}(a) has been performed for intermediate coupling $\varepsilon_{ls}$, and we expect that the effect of pressure is even stronger for smaller $\varepsilon_{ls}$ \cite{pham13}. 

To investigate the effect of pressure on the system dynamics in more detail, we plot in Fig. \ref{fig:P}(b) the vibrational spectrum of the confined liquid. The spectrum is calculated from the velocity autocorrelation function as
\begin{equation}
 S(\omega) \propto \sum_{i\in l} \int_{-\infty}^{\infty} dt e^{i\omega t} \langle v_i(t) v_i(0) \rangle. 
\end{equation}
For a harmonic solid, the spectrum $S(\omega)$ coincides with the phonon density of states and therefore provides direct information of atomic vibrations. For a liquid, the spectrum $S(\omega)$ contains, on the other hand, (indirect) information of both atomic vibrations and atomic diffusion \cite{march}. The contribution of diffusion to $S(\omega)$ is apparent from the fact that $S(\omega=0)$ is proportional to the diffusion constant in a bulk liquid \cite{march}. 

Figure \ref{fig:P}(b) shows at low pressure, vibrational spectrum is dominated by the low frequencies $\lesssim 2$ THz. The dominance of low frequencies highlights the ease of diffusion at low pressure and the related suppression of high-frequency vibrations. We believe that this suppression of high-frequency collective vibrations is intimately linked to the suppression of thermal conductance at high frequencies. When the pressure is increased to $100$ MPa, on the other hand, liquid-liquid interactions become stronger, which hinders diffusion and boosts high-frequency vibrations. This broadening of the spectrum enables more efficient energy transfer between the liquid and the solid at high frequencies. It is notable that the high-pressure spectrum exhibits a maximum at a non-zero frequency, typical of non-confined liquids \cite{marti01}. 

\section{Effect of liquid mass}
\label{app:E}
\begin{figure}
 \begin{center}
  \includegraphics[width=.99\columnwidth]{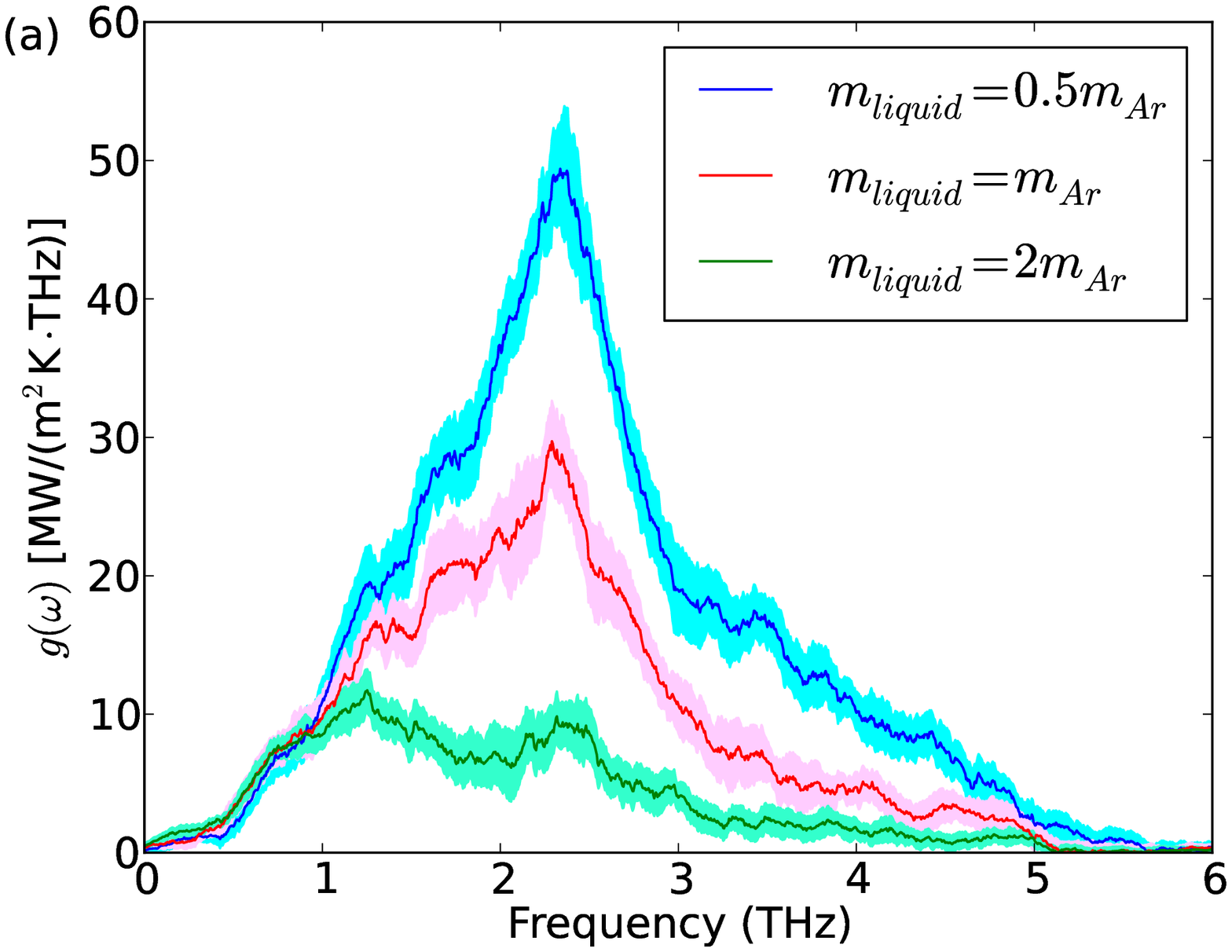} 
  \includegraphics[width=.99\columnwidth]{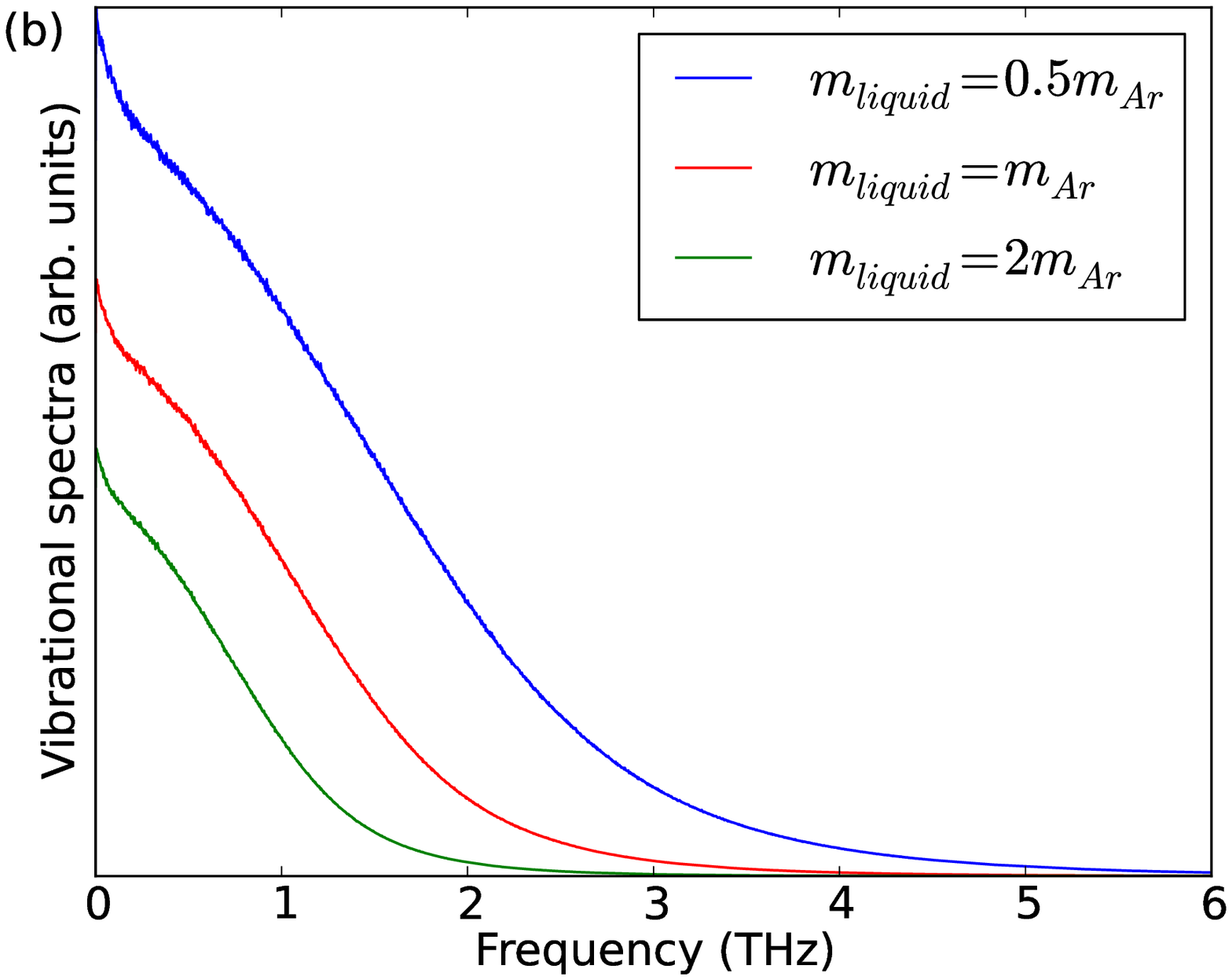}
\caption{(a) Spectral heat current $q(\omega)$ at the left liquid-solid interface for different liquid atom masses $m_{liquid}$. (b) Vibrational spectrum in the liquid in the three cases. While the calculation of vibrational density of states in the solid would require normalizing $S(\omega)$ by the atomic mass, we have not carried out this normalization for the liquid in order to highlight the effect of liquid mass on diffusion constant [proportional to $S(\omega=0)$].}
 \label{fig:mass}
 \end{center}
\end{figure}

The effect of the mass-mismatch of the liquid and solid atoms on the interfacial conductance has been investigated by Ge and Chen \cite{ge13b}. The authors noticed that increasing the solid-liquid mass ratio $m_{\textrm{Ar}}/m_{liquid}$ typically increases the conductance by enabling larger overlap in the vibrational spectra. By determining spectral heat currents, we can extend their analysis by investigating, which vibrational frequencies contribute to the enhanced conductance.

Figure \ref{fig:mass}(a) shows the spectral conductance for different liquid masses $m_{liquid}$. The mass of the solid atoms is $m_{Ar}$. Figure shows that at low frequencies $\lesssim 1$ THz, changing $m_{liquid}$ does not change the spectral conductance. This insensitivity at low frequencies (long wavelengths) suggests the inapplicability of the acoustic mismatch model (AMM) \cite{swartz89} on heat transfer at a classical liquid-solid interface. AMM predicts, after all, that the resistance is proportional to the mismatch in the acoustic impedances $Z_i=\rho_i c_i$, where $\rho$ and $c$ are the mass density and phonon velocity. No such dependence on density can be observed in Fig. \ref{fig:mass}(a) at low frequencies, where the long-wavelength approximation employed in AMM is valid.
 
At frequencies larger than 1 THz, on the other hand, the effect of liquid atom mass on the spectral conductance is strong. For small mass $m_{liquid}=0.5m_{Ar}$, the spectral conductance exhibits a very strong peak at $\sim$ 2.3 THz. When the liquid mass is larger than the solid atom mass ($m_{liquid}=2m_{Ar}$), the conductance distribution is, on the other hand, relatively flat, and the contribution of high-frequency modes ($\gtrsim 3$ THz) is small.

To further analyze the effect of liquid mass, we show in Fig. \ref{fig:mass}(b) the vibrational spectra for the three mass values. First, one can notice that reducing the mass enhances the diffusion, visible in the zero-frequency limit in the spectrum (proportional to the diffusion constant). This is intuitively expected, because the root-mean-square velocities of the atoms are inversely proportional to the square root of the mass and higher velocities are expected to lead to stronger diffusion. The case of low pressure (Fig. \ref{fig:P}) shows, however, that the diffusivity of liquid atoms does not play as important a role in liquid-solid heat transfer as the presence of high-frequency vibrations. Figure \ref{fig:mass}(b) shows that reducing the liquid mass notably broadens the spectrum and enables high-frequency vibrations in the liquid. Similarly to the case of high pressure (Fig. \ref{fig:P}), this broadening is accompanied by increased conductance.

\end{document}